\documentclass[
11pt,%
tightenlines,%
twoside,%
onecolumn,%
nofloats,%
nobibnotes,%
nofootinbib,%
superscriptaddress,%
noshowpacs,%
centertags]%
{revtex4}

\usepackage{ljm}
\usepackage[]{algorithm2e}

\usepackage{makecell}
 \renewcommand{\textit}[1]{
    $#1$
}

\usepackage{float}

\graphicspath{{./res/}}

\begin{document}

\titlerunning{Genetic Algorithm for Searching SQF pulse sequences for qubit control}

\authorrunning{Bastrakova, Kulandin, Laptyeva, Vozhakov, Liniov}

\title{Genetic Algorithm for Searching Bipolar Single-Flux-Quantum Pulse Sequences for Qubit Control}

\author{\firstname{M.~V.}~\surname{Bastrakova}}
\email[E-mail: ]{mailto:bastrakova@phys.unn.ru}
\affiliation{Lobachevsky State University of Nizhny Novgorod, 603022, Nizhny Novgorod, Prospekt Gagarina, 23}
\affiliation{Russian Quantum Center, 121205, Moscow, Russia}
\author{\firstname{D.~S.}~\surname{Kulandin}}
\email[E-mail: ]{mailto:kulandin08@gmail.com}
\affiliation{Lobachevsky State University of Nizhny Novgorod, 603022, Nizhny Novgorod, Prospekt Gagarina, 23}
\author{\firstname{T.}~\surname{Laptyeva}}
\email[E-mail: ]{mailto:tatyana.lapteva@itmm.unn.ru}
\affiliation{Lobachevsky State University of Nizhny Novgorod, 603022, Nizhny Novgorod, Prospekt Gagarina, 23}
\author{\firstname{V.~A.}~\surname{Vozhakov}}
\email[E-mail: ]{mailto:sevozh@yandex.ru}
\affiliation{Lomonosov Moscow State University, Skobeltsyn Institute of Nuclear Physics, 119991, Moscow}
\author{\firstname{A.~V.}~\surname{Liniov}}
\email[E-mail: ]{mailto:alin@unn.ru}
\affiliation{Lobachevsky State University of Nizhny Novgorod, 603022, Nizhny Novgorod, Prospekt Gagarina, 23}

\firstcollaboration{(Submitted by A.~V.~Liniov) }

\received{September 1, 2022}

\begin{abstract}
Nowadays most of superconducting quantum processors use charge qubits of a transmon type. They require implementation of energy efficient qubit state control scheme. A promising approach is the use of superconducting digital circuits operating with single-flux-quantum (SFQ) pulses. The duration of SFQ pulse control sequence is typically larger than that of conventional microwave drive pulses but its length can be optimized for the system with known parameters. Here we introduce a genetic algorithm for unipolar or bipolar SFQ control sequence search that minimize qubit state leakage from the computational subspace. The algorithm is also able to find a solution in the form of a repeating subsequence in order to save memory on the control chip. Its parallel implementation can find the appropriate sequence for arbitrary system parameters from a practical range in a reasonable time. The algorithm is illustrated by the example of the rotation gate  around the axis by an angle $\pi/2$ with fidelity over 99.99\%. In this paper, we present the results for a single-qubit system, but in the future we will apply the developed approach to study a system of two qubits.
\end{abstract}


\maketitle

\section{Introduction}

In recent years, there has been noticeable progress in the creation of quantum registers based on  Josephson junctions of various types \cite{Krantz2019,Kjaergaard2020,VozhakovUFN2022}: phase, flux, charge, etc. The classification of superconducting qubit types is based on the dominance of certain energy parameters, reminding of the classification of atoms in the periodic table. The most promising are the transmon qubits [3], in which the Josephson energy significantly exceeds the electrostatic one. Transmons are used as the basic units of modern quantum computers of the main research teams in the field of superconducting quantum computers: Google, IBM, Rigetti and others. They are characterized by high coherence, scalability, fast control. These qubits were developed in \cite{Koch2007} and their design scheme is a micrometre superconducting circuit consisting of a pair of paralleled capacitively-coupled Josephson junctions. This kind of circuit topology makes it possible to produce a strong localization of the junction phase and significantly reduce the effect of the charge noise in the system, considering the lower levels in the spectrum of the anharmonic Josephson oscillator as the qubit basic states.

The traditional method of qubit state control is the single-sideband modulation of a microwave carrier tone. It has been experimentally demonstrated that by fine-tuning the amplitudes of control in-phase and quadrature pulses, arbitrary rotation on the Bloch sphere can be performed. Thus, it is possible to implement any single-qubit operation. It is important to note that the weak transmon anharmonicity imposes a restriction on the duration and amplitude of the microwave pulses \cite{Koch2007}, which thereby limits the speed of the transmon registers. The duration of single-qubit operations with fidelity 99.99\% is 6-12 ns for DRAG technology~\cite{Werninghaus2021}. In this regard, various approaches are currently being discussed to avoid the use of high-frequency electronics for the implementation of computational algorithms. One of them is the use of digital devices of superconducting electronics \cite{Fedorov2007,Soloviev2014,Klenov2017,Bastrakova2020,Bastrakova2022} using sequences of control pulses with a wide spectrum - Single Flux Quantum (SFQ) pulses. This method has successfully proven the possibility to implement the single-qubit operations with regular sequences~\cite{McDermott2014} and optimized pulse (SCALLOPS) sequences~\cite{LiMcDermottVavilov2019} of SFQ pulses. It was possible to demonstrate the implementation of short (12-20 ns) single-qubit operations with the same fidelity as in microwave technology.

In this paper, we have shown that it is possible to accelerate quantum operations on a transmon qubit can by using bipolar sequences of the SFQ pulses. We have proposed an original genetic algorithm for searching for the bipolar sequences aimed at minimizing leakage from the computational qubit subspace (ground and first excited states). Further, we modified the algorithm aiming to look for sequences that can be constructed from smaller subsequences that are repeated certain amount of times. This idea can significantly reduce the memory size used by the SFQ pulses generator.  We show that the proposed optimization process of the bipolar SFQ sequences can be executed for a wide range of circuit parameters and the operation time is reduced to 4-10 ns, which is half as much compared to the known SCALLOP sequences~\cite{LiMcDermottVavilov2019}.

The rest of the paper is organized as follows.
Section 2 outlines the model of transmon qubit and its control with SFQ pulses.
The genetic algorithms for Searching SFQ pulse sequences are described in Section 3. Section 4 reports numerical results obtained for the model.
We summarized our findings and outline further perspectives in Section 5.

\section{Physical model of the transmon qubit}

We are considering an ordinary transmon qubit~\cite{Koch2007}. This qubit is connected via capacitance  $C_ñ$   to two SFQ generators - time-dependent voltage sources $\pm V(t)$. The connection scheme is shown in Figure~1. The Hamiltonian of the transmon can be represented as:
\begin{eqnarray}
\hat{H}_q = \frac{\hat{Q}^2}{2C} + E_J cos\hat{\phi},
\label{eq1}
\end{eqnarray}
where $C=C_c+C_q$, $\hat{Q}$ and $\hat{\phi}$ are the qubit charge and phase operators. Transmon is a standard two-contact SQUID, in which the Josephson energy $E_J$ significantly exceeds the electrostatic energy $E_C$ ($E_J = I_c \Phi_0/2 \pi$, $E_C=e^2/2C_q$, where $I_c$ is the critical current, $C_q$ is the capacitance of the Josephson junctions, $\Phi_0$ is the magnetic flux quantum).  

The interaction between the transmon and the DC-to-SFQ converter is described by the expression:
\begin{equation}
    \hat{H}_{SFQ}=\frac{C_ñ}{C}V(t) \hat{Q},
\label{eq2}
\end{equation}
\begin{figure}
\center
\includegraphics[width=0.5\columnwidth]{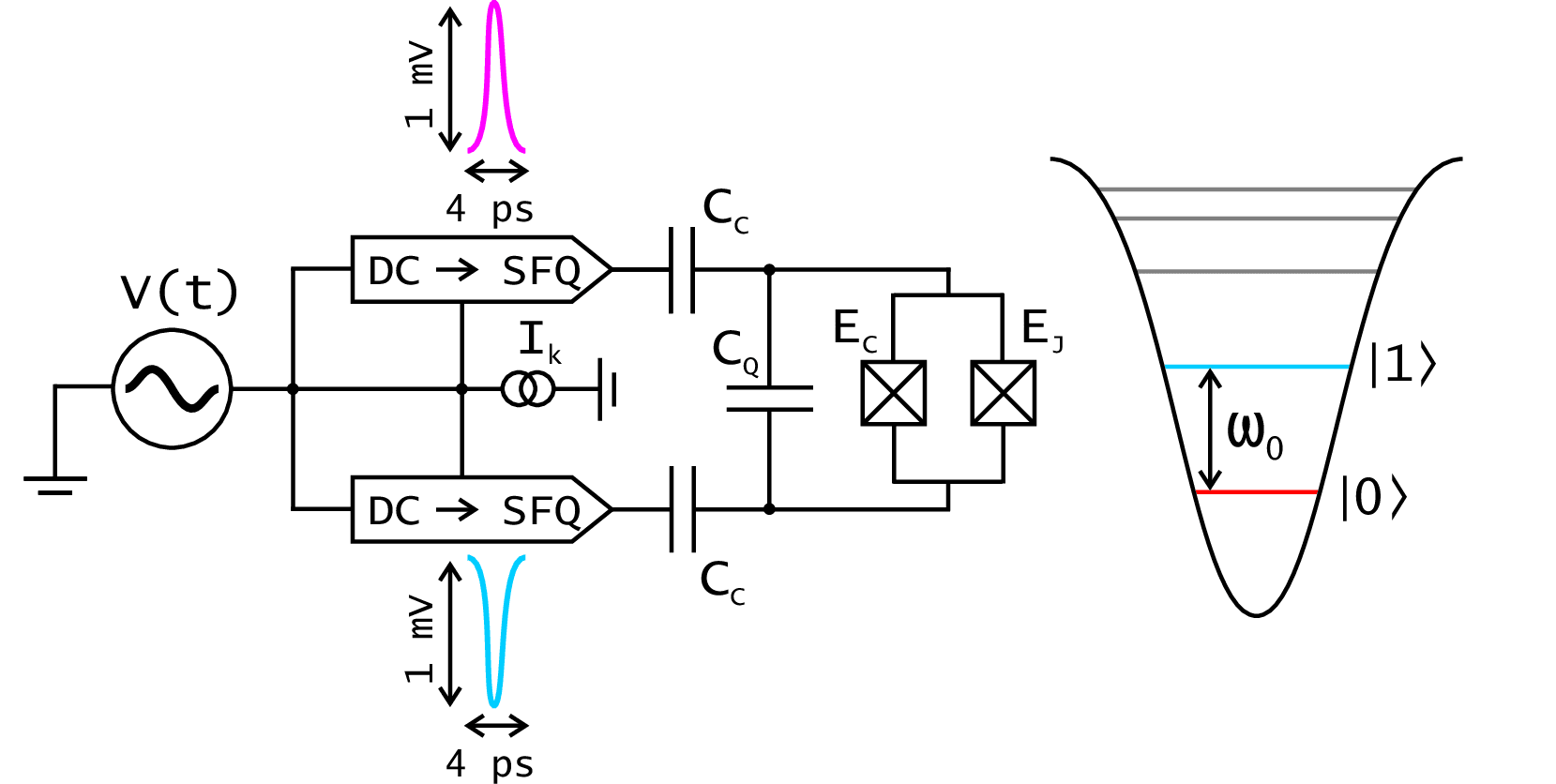}
\caption{\label{fig:transmon} Control scheme design and transmon energy spectrum}
\end{figure}
where SFQ pulse is determined by the magnetic flux quantum $\displaystyle \int^{\tau}_{0}V(t)dt = \Phi_0$ with duration $\tau$. The Schrodinger equation with the Hamiltonian in the form of~(\ref{eq1}) is solved exactly in terms of Mathieu functions for any ratio $E_J/E_C$, and the energy spectrum is characterized by strong dispersion \cite{Koch2007}. 

When the inequality $E_J \gg E_C$  is fulfilled, the dispersion – the dependence of the energy levels on the charge  – becomes "straightened". As a consequence, this leads to a strong decrease in the charge fluctuations influence on the qubit state. In addition, for $E_J \gg E_C$ case  the phase is highly localized, so the expression (\ref{eq1}) can be decomposed: 
$\cos \hat{\phi} = 1 - \frac{\hat{\phi}^2}{2!} + \frac{\hat{\phi}^4}{4!} + ...$. In terms of the usual raising  $\hat{a}^{\dagger}$  and lowering $\hat{a}$ operators, we can parametrize the total Hamiltonian of the qubit $\hat{H}=\hat{H}_q+\hat{H}_{SFQ}$ within the framework of the Bose-Hubbard model as: 
\begin{equation}
    \hat{H}= \hbar \omega_{0}\hat{a}^\dagger\hat{a} - \frac{\hbar \alpha}{2}  \hat{a}^\dagger\hat{a}\left(\hat{a}^\dagger\hat{a} -1\right) + \hbar \varepsilon (t)\left(\hat{a}+\hat{a}^\dagger\right), \label{eq3}
\end{equation}
where $\displaystyle \omega_{0}=\sqrt{8E_{J}E_{C}}$ is the qubit frequency (the transition frequency between the ground  $\left|0\right>$ and first excited state $\left|1\right>$), $ \displaystyle \alpha = E_C / 12$ is the qubit nonlinearity and $\displaystyle \varepsilon(t)= \frac{C_{C}V(t)}{2} \sqrt{\frac{\omega_{0}}{2C_Q}}$ is the amplitude of the SFQ pulse. A single SFQ pulse induces a discrete small rotation by the angle $\Delta \theta = C_c \Phi_0 \sqrt{\frac{\omega_0}{2C_q }}$  on the Bloch sphere corresponding to the qubit state change. The bipolar sequence of $M$ pulses, respectively, rotates the initial state of the qubit $\left|\alpha\right>$ by an angle  $\theta=\Delta\theta M$ according to the time-dependent Schrodinger equation:
\begin{equation}
    i\hbar \frac{\partial}{\partial t } |\psi(t)\rangle = \hat{H} |\psi(t)\rangle , \label{eq4}
\end{equation}
with the unitary evolution operator $U_g$: $|\psi(t)\rangle = U_{g} |\alpha\rangle$. Let's take, as an example, an ideal qubit operation - rotation around the $y$ axis. It's operator takes the form of $Y_{\pi/2}$: $U_{id} = e^{i \theta \sigma_y/2}$. Note, there is a nonzero probability of the leakage from the main qubit states $|0\rangle$, $|1\rangle$ to the overlying states $|2\rangle$, $|3\rangle$ etc. In this case, the fidelity of a quantum operation can be defined \cite{Bowdrey2002} as
\begin{equation}
    \langle F \rangle = \frac{1}{6}\sum_{|\alpha \rangle \in \nu} \left| \left<\alpha|U^{\dagger}_{g}U_{id}|\alpha\right>\right|^{2},\label{eq5}
\end{equation}
where the summation runs over the six states aligned along the cardinal directions of the Bloch sphere $|\alpha \rangle$ = $ \left\{|x_{\pm}\rangle = \dfrac{|0\rangle \pm |1\rangle}{\sqrt{2}}, \quad |y_{\pm}\rangle = \dfrac{|0\rangle\pm i |1\rangle}{\sqrt{2}}, \quad  |z_{+}\rangle = |0\rangle, \quad |z_{-}\rangle  = |1\rangle\right\}$. 
\section{Bipolar sequence search algorithms}

The work  ~\cite{LiMcDermottVavilov2019} presents qubit driving with so-called scalable leakage optimized pulse sequences (SCALLOPS)
which utilize more than one pulse during the qubit oscillation period, $2 \pi / {\omega}_0$.
This allows to make a $\pi / 2$ rotation for transmon qubits with their typical frequencies ${\omega}_0 / 2 \pi \approx$ 5 GHz at gate time above 12 ns.
The paper ~\cite{Vozhakov} extends the SCALLOP approach by introducing the pulses with negative polarity into the control sequences.
Bipolar SCALLOP sequences are optimized with the genetic-like algorithm aimed at the minimization of leakage.
In this paper, we describe two genetic algorithms that search for a bipolar control sequence with minimal leakage.
The first one searches for the best sequence of a given length.
The second searches for the optimal subsequence of a given length, the repetition of which a certain number of times (we consider from 1 to 35 repetitions) gives the minimum leakage (see ``Score Current Population'' of the algorithm).
Both algorithms are different only in the way the fitness function is calculated. Their pseudo-code is presented in Algorithm~\ref{algo} and includes the following steps.
\begin{enumerate}
\item Initialization

The genetic algorithm has a number of parameters: the probability of crossing $P_c$, the probability of mutation $P_m$, the population size $N$ and the number of generations $MaxIterations$.
We use the values selected based on the results of practical runs: $P_c = 0.8$, $P_m = 0.8$, $N = 2 * SequenceLength + 1$, $MaxIterations = 500$.
$SequenceLength$ is an input parameter.

The individuals are stored as arrays of integers with $SequenceLength$ elements, each element can store the values \{-1, 0, 1\} to represent a pulse with negative polarity ``-1'', an absence of pulse ``0'' and a pulse with positive polarity ``1'', respectively. At this stage, memory is allocated to store the current and next populations.

\item Generate Initial Population

We take a harmonic signal on the qubit frequency ${\omega}_0$ with the amplitude of the exciting field ${\varepsilon}_0$.
Then we choose a threshold value $\varepsilon_{th} \leq {\varepsilon}_0$ and place SFQ pulses at the extreme points of the drive frequency, $2\pi/\omega_{g}$, when the harmonic signal exceeds the threshold.
The pulse polarity is defined by the sign of the harmonic signal.
Depending on the values ${\omega}_0$, ${\omega}_g$, by choosing different $\varepsilon_{th}$, we can obtain about ${\omega}_g / 2 {\omega}_0$ such sequences.
The initial population includes a sequence with an approximately equal number of zero and non-zero elements and $2*N$ of its variations obtained by changing one of its elements to other possible values (for example, for an element with a value of ``-1'', the possible values are ``0'' and ``1'').

\item Main Algorithm Loop

The main loop of the algorithm performs a maximum of $MaxIterations = 500$ iterations, but can be terminated at any time when an individual is found that satisfies the conditions for the accuracy of the rotation angle $\Delta\theta$ and fidelity defined by the expression~(\ref{eq5}).

  \begin{enumerate} 
  \item Score Current Population

As the value of the individual's fitness function, a pair is used (accuracy of the target rotation angle, fidelity). When ordering the elements, the first values are checked and compared first, and in the case when both elements have a rotation angle error less than ${10}^{-4}$ - the second.
A variant of the algorithm that searches for the best subsequence of a given length calculates the accuracy of the rotation angle, $\Delta\theta$, and fidelity, $F$, for a different number of its repetitions from 1 to ${Max}_{rep}$. ${Max}_{rep}$ is an additional parameter of the algorithm. For example, it can be chosen in such a way that the maximum sequence length does not exceed some boundary value that specifies the maximum operation execution time.

  \item Selection/Crossover/Mutation Loop

The genetic algorithm uses three main types of rules several times ($L$ in our implementation) at each step to create the next generation from the current population.

  \begin{enumerate} 
    \item Select Parents

3 individuals are randomly selected from the population. 2 of them with the best fitness function values become parents.

    \item Combine Parents to Form Children

One Point Crossover is used. A random crossover point is selected and the tails of its two parents are swapped to get new offspring.

    \item Mutate Children

An analogue of Bit Flip Mutation is used - we select one random element and change it to one of the two other possible values.

    \item Add Children to Next Population
   \end{enumerate} 
   \item Adjust New Population

Two individuals of the new generation with the worst fitness function values are replaced by two individuals of the previous generation with the best.

   \end{enumerate} 
\item Save Results

The best sequence, the values of the obtained rotation angle and fidelity, and the running time of the algorithm are saved.

\end{enumerate}

More than 90\% of the algorithm execution time is occupied by the calculation of the individual fitness function value, which is performed at the ``Score Current Population'' step of the algorithm. Since these computations are independent, they can be parallelized using any parallel programming technology. We use OpenMP, but did not perform possible code optimizations for systems with the NUMA architecture, since the performance of the current implementation is sufficient to solve all the problems considered.

\begin{algorithm}
\tcp{Length of the control sequence/subsequence is predefined}
\tcp{Algorithm parameter values are selected based on practical use}
Initialization\;
const L = Length of the Control Sequence\;
const N = 2 * L + 1\;
Generate Initial Population\;
\For{i = 0 \KwTo MaxIterations}{
  Score Current Population\;
  \For{i = 0 \KwTo L}{ 
    Select Parents\;
    Combine Parents to Form Children\;
    Mutate Children\;
    Add Children to Next Population\;
  }
  Adjust New Population\;
}
Save Results\;
\caption{\label{algo}Genetic algorithm for searching SFQ pulse sequences for qubit control}
\end{algorithm}

The unipolar version of the algorithm is obtained by changing the set of possible values of the sequence elements from \{-1, 0, 1\} to \{0, 1\} with the appropriate changes at the ``Mutate Children'' step.
The unipolar version was used only to check the correctness of the algorithm. When run with the input given in article~\cite{LiMcDermottVavilov2019}, it found sequences with comparable angle rotation and fidelity values. 
In the section with experimental data, only results of the bipolar algorithm runs are presented.

\section{SFQ pulse sequences search results}

\begin{figure}[H]
\center
\includegraphics[width=0.6\columnwidth]{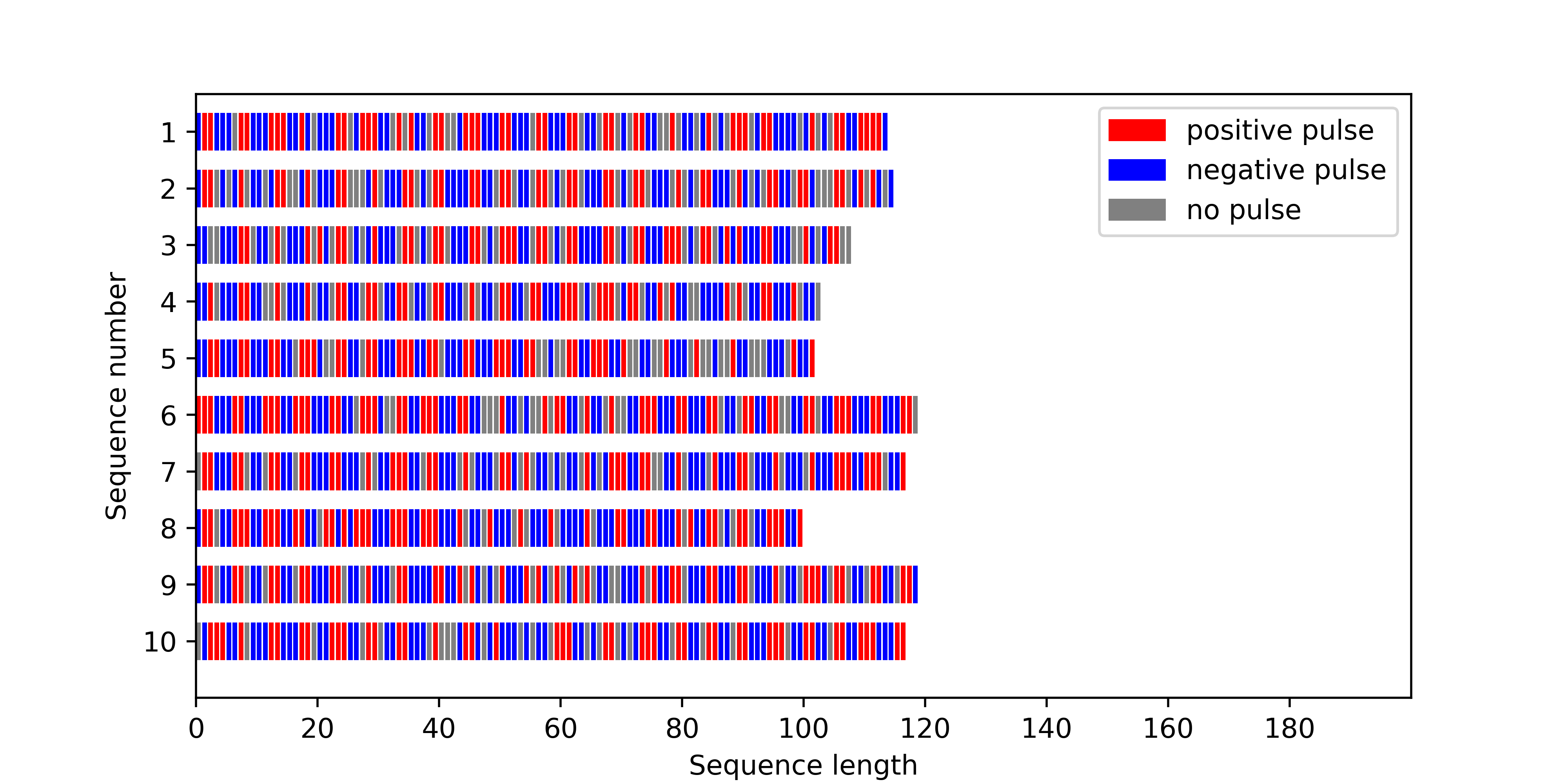}
\caption{\label{fig:sequences} SFQ pulse control sequences providing ${Y}_{\pi / 2}$ gate for the first 10 parameter sets.}
\end{figure}

\begin{figure}[H]
\center
\includegraphics[width=0.5\columnwidth]{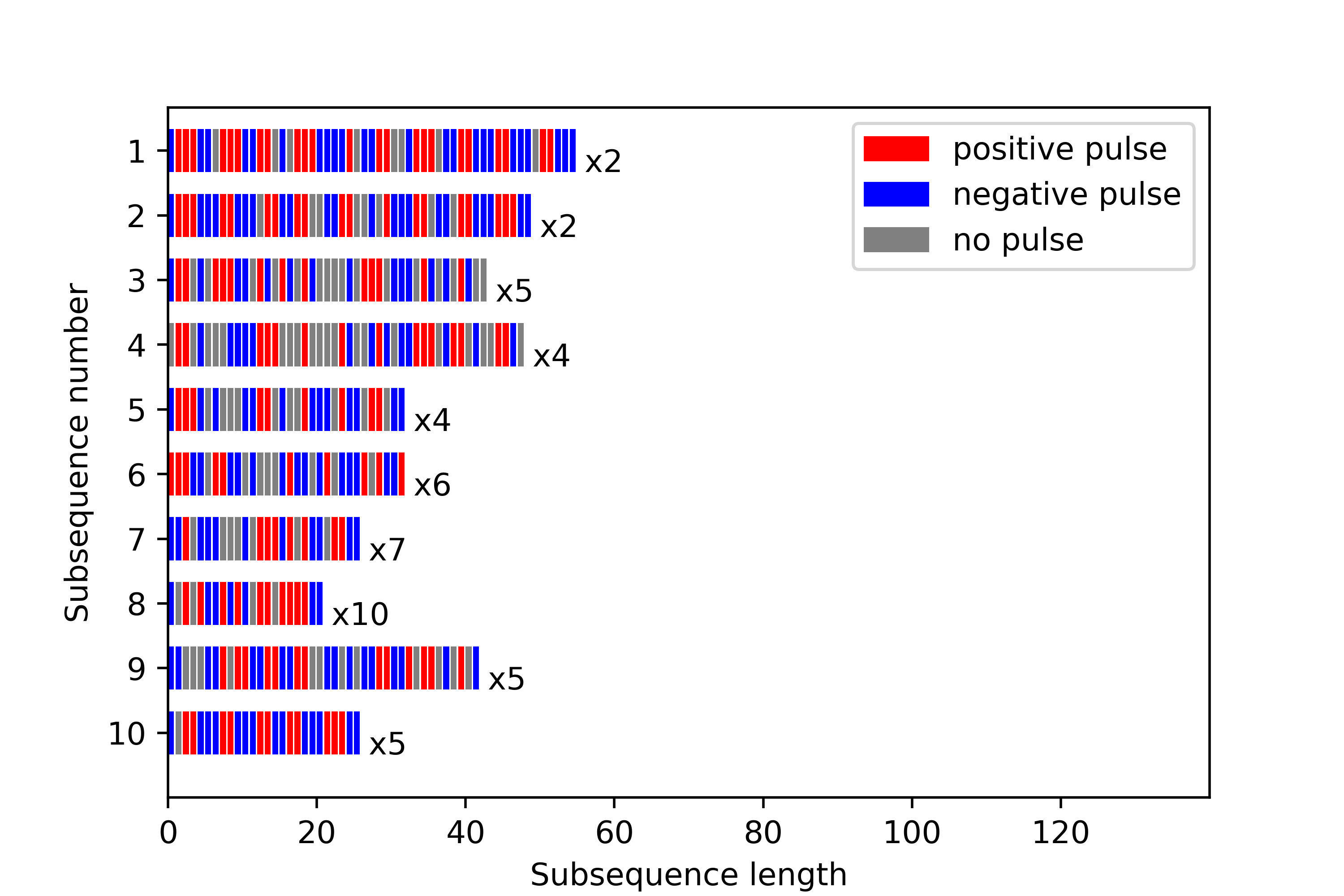}
\caption{\label{fig:subsequences} SFQ pulse control subsequences providing ${Y}_{\pi / 2}$ gate for the first 10 parameter sets.}
\end{figure}

\begin{figure}[H]
\center
\includegraphics[width=0.7\columnwidth]{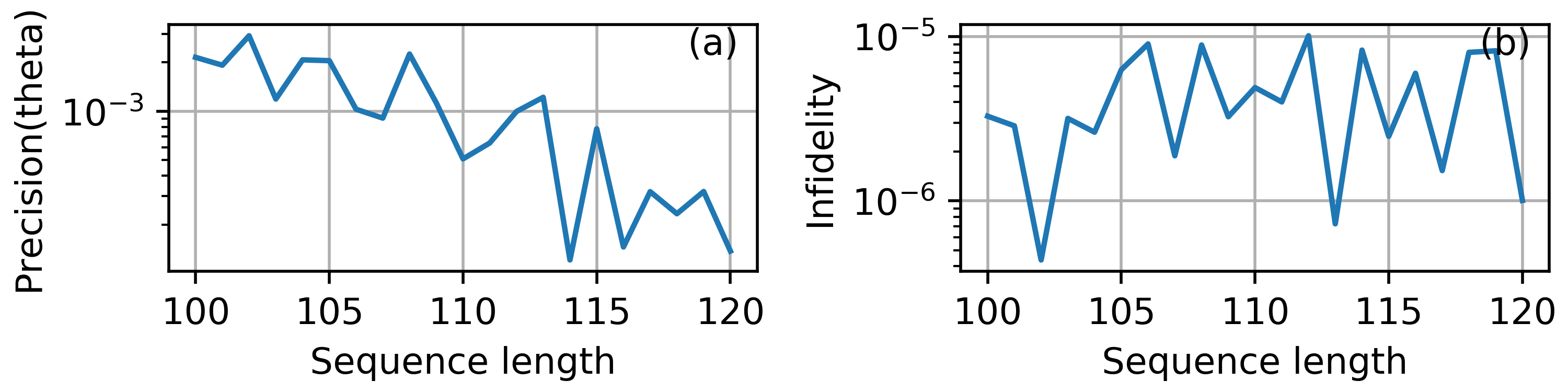}
\caption{\label{fig:sequenceslength} The dependence of the (a)~rotation accuracy and (b)~infidelity on the length of the sequence for the set of parameters N 1 in Table~1.}
\end{figure}

\begin{figure}[H]
\center
\includegraphics[width=0.7\columnwidth]{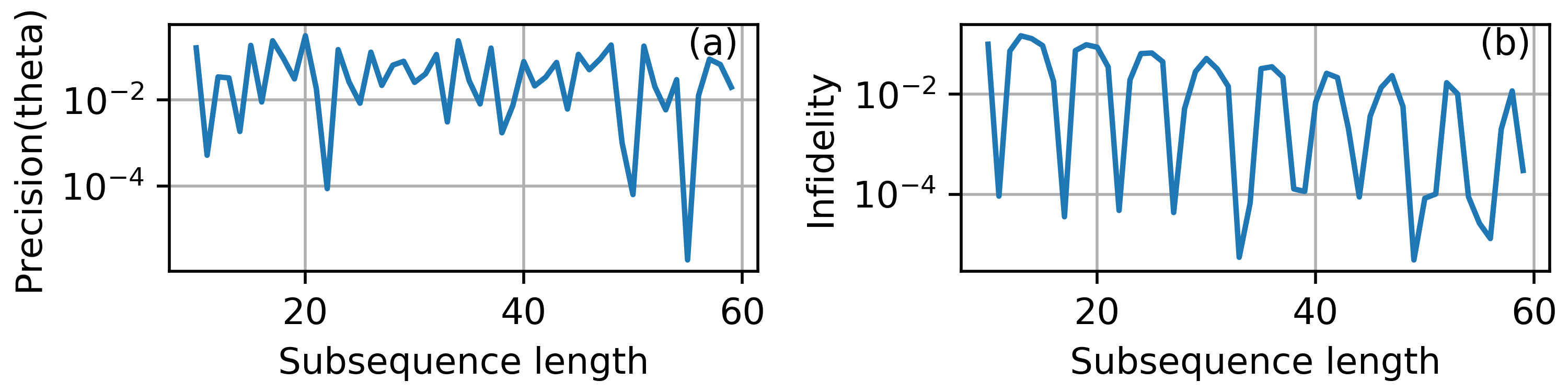}
\caption{\label{fig:subsequenceslength} The dependence of the (a)~rotation accuracy and (b)~infidelity on the length of the subsequence for the set of parameters N 1 in Table~2.}
\end{figure}

We have searched for SFQ pulse control sequences for the typical sets of the system parameters used in~\cite{LiMcDermottVavilov2019}: $\omega_{0} / 2\pi$ = \{4.54643, \dots, 5.48906\} GHz, $\omega_{g} / 2\pi$ = 25 GHz, $\Delta \theta$ = 0.032 rad.
The found sequences and their parameters for the sequence search algorithm are shown in Table~\ref{tbl:tbl1} and Figure~\ref{fig:sequences}. The results of the subsequence search algorithm are shown in Table~\ref{tbl:tbl2} and Figure~\ref{fig:subsequences}. The results of calculations show that the the operation time is reduced by about two times relative to the work~\cite{LiMcDermottVavilov2019}.
When searching for sequences, for every $N$ one can find an individual with a rotation accuracy better than ${10}^{-5}$. For each $\hat{H}$, Eq.~(\ref{eq3}), the Table~\ref{tbl:tbl1} contains data for the sequence, which shows the minimum value of infidelity, $1 - F$. 
The use of subsequences does not always provide an acceptable rotation accuracy. The Table~\ref{tbl:tbl2} for each $N$ shows the data of the subsequence of the minimum length with the rotation angle error less than ${10}^{-5}$ and infidelity less than ${10}^{-4}$.
Typical dependences of the rotation accuracy and fidelity for the best found solution on the  sequence/subsequence length is shown on Figure~\ref{fig:sequenceslength} and Figure~\ref{fig:subsequenceslength}.

\begin{table}[H]
    \setcaptionmargin{0mm}
    \onelinecaptionstrue
    \captionstyle{flushleft}
    \caption{SFQ pulse control sequences providing ${Y}_{\pi / 2}$ gate obtained using the sequence search algorithm for various transmon parameters under a bipolar SFQ drive.}
    \label{tbl:tbl1}
    \bigskip
    \centering
    \begin{tabular}{|c|c|c|c|c|c|c|c|}
        \hline
        N & $\Delta \theta$, rad & ${\omega}_0$ & Sequence length & \makecell{Operation \\ duration, ns} & Precision($\theta$), ${10}^{-7}$ & $1 -  F, {10}^{-5} $ & \makecell{Algorithm\\execution time, s}\\
        \hline
 1 & 0.032 & 4.54643 & 114 & 4.52 & 9.25 & 9.51 & 21.74 \\
        \hline
 2 & 0.032 & 4.59251 & 115 & 4.56 &  3.48 & 1.35 & 21.57\\
        \hline
 3 & 0.032 & 4.6305   & 108 & 4.28 & 12.46 & 0.11 & 20.59\\
        \hline
 4 & 0.032 & 4.652    & 103 & 4.08 & 2.82 & 1.83 & 20.69\\
        \hline
 5 & 0.032 & 4.68842  & 102 & 4.04 & 50.24 & 8.48 & 19.77\\
        \hline
 6 & 0.032 & 4.73047  & 119 & 4.72 & 63.40 & 8.20 & 22.63\\
        \hline
 7 & 0.032 & 4.76289  & 117 & 4.64 & 2.74 & 1.37 & 22.11\\
        \hline
 8 & 0.032 & 4.78802  & 100 & 3.96 & 2.40 & 5.73 & 19.42\\
        \hline
 9 & 0.032 & 4.80851  & 119 & 4.72 & 0.98 & 0.17 & 21.71\\
       \hline
10 & 0.032 & 4.87898  & 117 & 4.64 & 0.24 & 2.45 & 22.46\\
       \hline
11 & 0.032 & 4.89201  & 104 & 4.12 & 3.62 & 2.33 & 21.49\\
       \hline
12 & 0.032 & 4.90296  & 116 & 4.60 & 1.36 & 5.16 & 22.95\\
       \hline
13 & 0.032 & 5.18978  & 115 & 4.56 & 0.10 & 3.46 & 21.91\\
       \hline
14 & 0.032 & 5.20945  & 119 & 4.72 & 0.03 & 1.07 & 22.71\\
       \hline
15 & 0.032 & 5.28923  & 113 & 4.48 & 1.66 & 3.87 & 22.60\\
       \hline
16 & 0.032 & 5.32036  & 111 & 4.40 & 2.48 & 2.49 & 22.60\\
       \hline
17 & 0.032 & 5.35835  & 114 & 4.52 & 0.60 & 1.25 & 22.83\\
       \hline
18 & 0.032 & 5.39307  & 120 & 4.76 & 2.36 & 1.09 & 25.44\\
       \hline
19 & 0.032 & 5.40655  & 114 & 4.52 & 3.38 & 1.60 & 24.17\\
       \hline
20 & 0.032 & 5.43571  & 114 & 4.52 & 6.30 & 1.85 & 22.43\\
       \hline
21 & 0.032 & 5.48906  & 120 & 4.76 & 13.64 & 0.47 & 24.17\\
       \hline
    \end{tabular}
\end{table}

\begin{table}[H]
    \setcaptionmargin{0mm}
    \onelinecaptionstrue
    \captionstyle{flushleft}
    \caption{SFQ pulse control sequences providing ${Y}_{\pi / 2}$ gate obtained using the subsequence search algorithm for various transmon parameters under a bipolar SFQ drive.}
    \label{tbl:tbl2}
    \bigskip
    \centering
    \begin{tabular}{|c|c|c|c|c|c|c|c|c|c|}
        \hline
        N & $\Delta \theta$, rad & ${\omega}_0$ & \makecell{Subsequence\\length} & \makecell{Number of\\repetitions} & \makecell{Sequence\\length} &   \makecell{Operation \\ duration, ns} & Precision($\theta$), ${10}^{-6}$ & $1 -  F, {10}^{-5} $ & \makecell{Algorithm\\execution time, s}\\
        \hline
1 & 0.032 & 4.54643 & 55 & 2 & 110 & 4.40 & 2.66 & 1.39 & 95.91\\
       \hline
2 & 0.032 & 4.59251 & 49 & 2 & 98 & 3.92 & 4.41 & 8.18 & 111.08\\
       \hline
3 & 0.032 & 4.6305 & 43 & 5 & 215 & 8.60 & 5.61 & 3.53 & 94.97\\
       \hline
4 & 0.032 & 4.652 & 48 & 4 & 192 & 7.68 & 5.88 & 5.42 & 115.72\\
       \hline
5 & 0.032 & 4.68842 & 32 & 4 & 128 & 5.12 & 4.94 & 7.22 & 100.74\\
       \hline
6 & 0.032 & 4.73047 & 32 & 6 & 192 & 7.68 & 5.63 & 1.60 & 108.61\\
       \hline
7 & 0.032 & 4.76289 & 26 & 7 & 182 & 7.28 & 6.28 & 7.61 & 121.70\\
       \hline
8 & 0.032 & 4.78802 & 21 & 10 & 210 & 8.40 & 6.16 & 0.91 & 131.35\\
       \hline
9 & 0.032 & 4.80851 & 42 & 5 & 210 & 8.40 & 0.88 & 0.59 & 123.20\\
       \hline
10 & 0.032 & 4.87898 & 26 & 5 & 130 & 5.20 & 2.45 & 5.41 & 118.44\\
       \hline
11 & 0.032 & 4.89201 & 36 & 8 & 288 & 11.52 & 15.20 & 1.41 & 98.14\\
       \hline
12 & 0.032 & 4.90296 & 36 & 6 & 216 & 8.64 & 0.73 & 1.21 & 100.85\\
       \hline
13 & 0.032 & 5.18978 & 38 & 3 & 114 & 4.56 & 8.02 & 2.42 & 84.70\\
       \hline
14 & 0.032 & 5.20945 & 19 & 8 & 152 & 6.08 & 0.23 & 1.82 & 82.96\\
       \hline
15 & 0.032 & 5.28923 & 43 & 4 & 172 & 6.88 & 2.87 & 2.52 & 98.04\\
       \hline
16 & 0.032 & 5.32036 & 19 & 8 & 152 & 6.08 & 5.44 & 2.38 & 89.25\\
       \hline
17 & 0.032 & 5.35835 & 55 & 2 & 110 & 4.40 & 6.65 & 5.63 & 105.87\\
       \hline
18 & 0.032 & 5.39307 & 28 & 5 & 140 & 5.60 & 4.08 & 1.78 & 108.71\\
       \hline
19 & 0.032 & 5.40655 & 42 & 3 & 126 & 5.04 & 1.51 & 9.71 & 122.01\\
       \hline
20 & 0.032 & 5.43571 & 51 & 4 & 204 & 8.16 & 8.87 & 0.54 & 97.72\\
       \hline
21 & 0.032 & 5.48906 & 37 & 3 & 111 & 4.44 & 8.87 & 0.94 & 110.17\\
       \hline
    \end{tabular}
\end{table}

Numerical experiments were performed on the Lobachevsky supercomputer~\cite{SCLobachevsky} at the Lobachevsky State University of Nizhny Novgorod.
We employed computing nodes with the following configuration: 2x Intel XeonSilver 4310T CPU (20 cores, 2.30 GHz), 64 GB RAM, OS CentOS 7.9. We use Intel C/C++ Compiler and Intel Math Kernel Library (MKL) from Intel OneAPI HPC Toolkit~\cite{oneAPI}.

When using the OpenMP technology for parallelizing calculations over sequences/subsequences, the acceleration of solving one problem on 20 physical cores was $\sim3.7$ when searching for sequences and $\sim2.7$ when searching for subsequences. The low efficiency of such parallelization is due to the very small amount of data used and frequent switching between parallel and serial parts of code.

However, another variant of parallelization is possible. Several computations for different sets of parameters, but the same lengths of sequences/subsequences, can be run on the same node in serial mode with binding each computation to a separate core. This calculation scheme can be implemented using the OpenMP technology, or starting programs through the numactl utility, or using the capabilities of a cluster management system (for example, slurm). When running 20 calculations in parallel on 20 physical cores, we obtained a parallelization efficiency of about 80\% when searching for sequences and 70\% when searching for subsequences.

\section{Conclusions \& Future Work}

We present the simulation of coherent control of the transmon qubits using irradiation with sequences of SFQ pulses of different polarities. We selected a set of input parameters (similar to those presented in \cite{LiMcDermottVavilov2019}) consisting of 21 qubit frequencies controlled by a single global clock at 25~GHz generator frequency. We have proposed the genetic algorithm for generating bipolar sequences. It is shown that our algorithm allows us to find sequences of SFQ pulses suitable to implement the required quantum single-qubit operation with the fidelity better than 99.99\% by minimizing leakage from the computational subspace of the qubit. We have demonstrated, using the example of the qubit state rotation operation $Y_{\pi/2}$, that the use of the proposed bipolar sequences more than double the speed of the quantum gate, compared with the results of~~\cite{LiMcDermottVavilov2019}. In addition, the algorithm is able to search for a solution in the form of a repeating subsequence, which saves memory on the control chip. Such a control method can be used to implement two-qubit gates, similar to the microwave cross-resonance technique~\cite{Chow2011}. At the same time, we believe that the genetic algorithm will not undergo significant changes. It will only be necessary to change the physical model of the system and the number of required sequences acting on each of the qubits. This is our further plans for the development of this work.

\begin{acknowledgments}
The analysis of superconducting circuits and calculation of dynamics was carried out with the support of the grant of the President of the Russian Federation No.MK-2740.2021.1.2 and RFBR No.20-07-00952. The development of the optimization algorithm is performed under support of the Scientific and Educational Center "Mathematics of Future Technologies" (agreement no. 075-02-2022-883). The M.B. acknowledges the Basis Foundation scholarship.

\end{acknowledgments}



\end{document}